\documentclass[aps,preprint,nofootinbib,superscriptaddress]{revtex4}

\usepackage{amssymb}

\usepackage{epsfig}
\usepackage[bookmarksnumbered,bookmarksopen,colorlinks,citecolor=blue,linkcolor=blue]{hyperref}

\begin{document}

\title{ Microscopic dynamics simulations of heavy-ion fusion reactions induced by neutron-rich nuclei }

\author{Ning Wang}
\email{wangning@gxnu.edu.cn}\affiliation{ Department of Physics,
Guangxi Normal University, Guilin 541004, People's Republic of
China }

\author{Li Ou}
\affiliation{ Department of Physics,
Guangxi Normal University, Guilin 541004, People's Republic of
China }

\author{Yingxun Zhang}
\affiliation{China Institute of Atomic Energy, Beijing 102413, People's Republic of
China}

\author{Zhuxia Li}
\affiliation{China Institute of Atomic Energy, Beijing 102413, People's Republic of
China}

\begin{abstract}

The heavy-ion fusion reactions induced by neutron-rich nuclei are investigated with the improved quantum molecular dynamics
(ImQMD) model. With a subtle consideration of the neutron skin thickness of nuclei and the symmetry potential, the stability of nuclei and the fusion excitation functions of heavy-ion fusion reactions $^{16}$O+$^{76}$Ge, $^{16}$O+$^{154}$Sm, $^{40}$Ca+$^{96}$Zr and $^{132}$Sn+$^{40}$Ca are systematically studied. The fusion cross sections of these reactions at energies around the Coulomb barrier can be well reproduced by using the ImQMD model. The corresponding slope parameter of the symmetry energy adopted in the calculations is $L \approx 78$ MeV and the surface energy coefficient is $g_{\rm sur}=18\pm 1.5$ MeVfm$^2$. In addition, it is found that the surface-symmetry term significantly influences the fusion cross sections of neutron-rich fusion systems. For sub-barrier fusion, the dynamical fluctuations in the densities of the reaction partners and the enhanced surface diffuseness at neck side result in the lowering of the fusion barrier.

\end{abstract}
\maketitle

\begin{center}
\textbf{I. INTRODUCTION}
\end{center}

The synthesis of super-heavy nuclei and heavy-ion fusion reactions induced by neutron-rich nuclei
have attracted much attention in recent years \cite{Hof00,Ogan10,Ada,Zhou11,Wang11,Sar,Timm}. The calculations of the fusion (capture) excitation functions are of significant importance for the study of the nuclear structure and test of the models. The fusion (capture) cross sections of heavy-ion reactions can be predicted with some static models \cite{prox,liumin06,Wang09,MWS,Zag08,Cap11,Hag99}, in which the nucleus-nucleus potential is calculated under frozen density approximation or simply described by using the Woods-Saxon potential and the fusion (capture) probability is then obtained based on the barrier penetration concept together with the calculated nucleus-nucleus potential. The influence of the microscopic effects on the fusion barrier are empirically described by the barrier distribution function or absorbed in the model parameters. To self-consistently consider the dynamical effects in the fusion reactions, some microscopical dynamics models, such as the time-dependent Hartree-Fock (TDHF) model \cite{Umar06,Umar12}, the improved quantum molecular dynamic (ImQMD) model \cite{ImQMD2002,ImQMD2004,ImQMD2010} and the Vlasov simulation plus imaginary times approach \cite{Bona94,Bona00,Bona97} have been developed. The nucleus-nucleus potential and the fusion cross sections at energies above the Coulomb barrier can be successfully described with the TDHF calculations based on the the Skyrme energy-density functional describing the interaction between nucleons. However, there still exists some difficulties for the TDHF theory to describe the tunnelling of the many-body wave-function. As a consequence, there is no sub-barrier fusion \cite{Sim14}. The fusion (capture) cross sections at sub-barrier energies are indirectly calculated with the obtained nucleus-nucleus potential from the density-constrained TDHF technique \cite{Umar06} together with the barrier penetration concept. To investigate the dynamical behavior of fusion reactions at energies around the Coulomb barrier, especially the reactions induced by neutron-rich nuclei which could provide some helpful information on the synthesis of super-heavy nuclei, the microscopic dynamics model should be further improved and the model parameters should be refined to properly describe the neutron-rich systems.

As a semi-classical microscopic dynamics model, the quantum molecular dynamics (QMD) model \cite{QMD} was proposed for simulating heavy-ion collisions (HICs) at intermediate and high energies. With great efforts to develop the QMD model, some different extended versions of the QMD model such as IQMD \cite{IQMD93,IQMD99}, CoMD \cite{constrain,Maru02}, ImQMD and UrQMD \cite{UrQMD} have been proposed in the literature. To extend the QMD model for the study of heavy-ion reactions at energies around the Coulomb barrier, the ImQMD model was proposed based the QMD framework with some modifications: (1) The standard Skyrme force is adopted for describing not only the bulk properties but also the surface properties of nuclei; (2) The phase-space occupation constraint method is used following the CoMD model; (3) The mass dependence of the wave-packet width is considered for a better description of the surface properties of finite nuclei and the fluctuations in reactions. (4) The momentum dependence of the nucleon-nucleon interaction according to the Skyrme force is involved for the HICs at intermediate energies \cite{Zhang14}. It is found that the ImQMD model is successfully applied on HICs at intermediate energies \cite{ImQMD05,ImQMD13,Zhang14} and heavy-ion fusion reactions between stable nuclei \cite{ImQMD2004,ImQMD2012,Wen13}. It is well known that the symmetry energy plays an important role on the structure of neutron-rich nuclei. The thickness of neutron skin of nuclei has been explored to be linearly correlated with the slope of symmetry energy and the isospin asymmetry $I=(N-Z)/A$ of nuclei. For fusion reactions induced by neutron-rich nuclei, it is expected that the neutron-skin thickness of neutron-rich nuclei affects the nucleus-nucleus potential and consequently the fusion cross sections. It is therefore interesting to test the model and to investigate the nuclear symmetry energy from the neutron-rich fusion reactions.

In this work, we will systematically investigate the fusion reactions induced by neutron-rich nuclei based on the ImQMD simulations. The influence of the model parameters especially the parameters related to the symmetry potential on the stability of neutron-rich nuclei and the fusion excitation functions will be studied. The structure of this paper is as follows: In sec. II, the mean-field and the initialization of the ImQMD model are introduced. In sec. III, the fusion cross sections of some reactions with neutron-rich nuclei will be presented. The influence of the surface-symmetry term and dynamical fluctuations will also be investigated. Finally a brief summary is given in Sec. IV.

\begin{center}
\noindent{\bf {II. THE IMPROVED QUANTUM MOLECULAR DYNAMICS MODEL}}\\
\end{center}

In this section we first briefly introduce the framework of the ImQMD model, then the initialization of the model will be introduced for the reader's convenience. Simultaneously, the stability of nuclei will be studied with the proposed model parameters.
\begin{center}
\textbf{A. Mean-field in the Model  }
\end{center}

In the ImQMD model, as in the original QMD model,  each
nucleon is represented by a coherent state of a Gaussian wave
packet. The density distribution function $\rho$ of a system reads
\begin{equation} \label{1}
\rho(\mathbf{r})=\sum_i{\frac{1}{(2\pi \sigma_r^2)^{3/2}}\exp
\left [-\frac{(\mathbf{r}-\mathbf{r}_i)^2}{2\sigma_r^2} \right ]},
\end{equation}
where $\sigma_r$ represents the spatial spread of the wave packet.
The propagation of nucleons is governed by the self-consistently generated mean field,
\begin{equation} \label{2}
\mathbf{\dot{r}}_i=\frac{\partial H}{\partial \mathbf{p}_i}, \; \;
\mathbf{\dot{p}}_i=-\frac{\partial H}{\partial \mathbf{r}_i},
\end{equation}
where $r_i$ and $p_i$ are the center of the $i$-th wave packet in
the coordinate and momentum space, respectively.  The Hamiltonian
$H$ consists of the kinetic energy
$T=\sum\limits_{i}\frac{\mathbf{p}_{i}^{2}}{2m}$ and the effective
interaction potential energy $U$:
\begin{equation} \label{3}
H=T+U.
\end{equation}
The effective interaction potential energy is written as the sum
of the nuclear interaction potential energy $U_{\rm
loc}=\int{V_{\rm loc}(\textbf{r})d\textbf{r}}$ and the Coulomb
interaction potential energy $U_{\rm Coul}$ which includes the
contribution of the direct and exchange terms,
\begin{equation}
U=U_{\rm loc}+U_{\rm Coul}.
\end{equation}
Where $V_{\rm loc}(r)$ is the potential energy density that is
obtained from the effective Skyrme interaction without the spin-orbit term:
\begin{equation}
V_{\rm
loc}=\frac{\alpha}{2}\frac{\rho^2}{\rho_0}+\frac{\beta}{\gamma+1}\frac{\rho^{\gamma+1}}{\rho_0^{\gamma}}+\frac{g_{\rm
sur}}{2\rho_0}(\nabla\rho)^2
+g_{\tau}\frac{\rho^{\eta+1}}{\rho_0^{\eta}}+\frac{C_s}{2\rho_0}[\rho^2-k_s(\nabla\rho)^2]\delta^2
\end{equation}
where $\delta=(\rho_n -\rho_p)/(\rho_n +\rho_p)$ is the isospin
asymmetry. The corresponding density dependence of nuclear symmetry energy in cold nuclear matter is expressed as
\begin{equation}
 E_{\rm sym}(\rho)= 13 \left(\frac{\rho }{\rho_0 }\right )^{2/3} + \frac{C_s}{2 }\left(\frac{\rho }{\rho_0 }\right ).
\end{equation}
Here, the first term describes the contribution of the kinetic energy. The slope parameter of the symmetry energy at the saturation density is given by
\begin{eqnarray}
L=3\rho_0 \left (\frac{\partial E_{\rm sym}}{\partial \rho} \right
)_{\rho=\rho_0}.
\end{eqnarray}

To describe the fermionic nature of the N-body system and to improve the stability of an individual nucleus, the modified Fermi constraint \cite{ImQMD2010} in which the total energy of the system at the next time step is simultaneously checked after performing the two-body elastic scattering in the phase-space occupation constraint method \cite{constrain} is adopted. In addition, considering the fact that the mean-field plays a dominant role in heavy-ion fusion reactions at energies around the Coulomb barrier, the collision term in the traditional QMD model is not involved in the present calculations in order to eliminate the uncertainty of the parameters in the collision term due to the uncertainty of the medium effect in nucleon-nucleon cross sections and the different methods to deal with the Pauli blocking.

In the previous works \cite{ImQMD2012,ImQMD13}, the parameter set IQ3 is proposed and tested for describing the heavy-ion fusion reactions and the multifragmentation process in intermediate energy collisions between stable nuclei. With an incompressibility coefficient of $K_{\infty} = 225$ MeV and a relatively small width of the wave-packets $\sigma_r =  \sigma_0 + \sigma_1 A^{1/3} $ in the coordinates, IQ3 can reasonably reproduce the fusion excitation function of $^{16}$O+$^{208}$Pb at energies above the Coulomb barrier and the charge distribution of $^{197}$Au+$^{197}$Au at Fermi energies. In this work, the parameters related to the surface properties of finite nuclei and the symmetry potential in IQ3 are refined for a better description of the heavy-ion fusion reactions induced by neutron-rich nuclei. It is known that the nuclear symmetry energy at saturation density is about $E_{\rm sym} (\rho_0)\approx 30$ MeV, which is helpful to constrain the values of the parameters $C_s\approx 34$ MeV.  Three sets of model parameters are listed in Table I.  The parameter set SkP* is generally determined based on the Skyrme force SkP \cite{skp} in which the parameters $g_\tau$, $\sigma_0$ and $\sigma_1$ are adjusted for an appropriate description of nuclear properties at ground state and the fusion reactions.  The corresponding value of the incompressibility coefficient of nuclear matter is $K_{\infty} = 195$ MeV for SkP*, and 225 MeV for both IQ3a and IQ3b.  According to Eq.(7), the slope parameter of the symmetry energy for the three sets of parameters is about $L\approx 78$ MeV, which locates in the region of $L=70\pm 15$ MeV predicted by the latest finite range droplet model \cite{FRDM12} and of $ 53<L<79$ MeV  extracted from the nuclear masses together with the semi-empirical connection between the symmetry energy coefficients of finite nuclei and the nuclear symmetry energy at reference densities \cite{Liu10}. In addition, the value of the surface  coefficient $g_{\rm sur}$ should also be investigated and constrained since the fusion barrier is closely related to the surface properties of nuclei. In Fig. 1, we show the distribution of the corresponding surface coefficient $g_{\rm sur}=\frac{1}{32}(9t_1-5t_2-4 x_2 t_2)\rho_0$ from 103 sets of Skyrme forces with  $K_{\infty} = 230 \pm 40$ MeV. Here, $t_1$, $t_2$ and $x_2$ are the parameters of the Skyrme forces. One sees that the surface coefficient adopted in different Skyrme forces has the value around $g_{\rm sur}=20 \pm 5$ MeVfm$^{2}$. Over the 103 sets of Skyrme forces, 16 sets have $g_{\rm sur}\approx 16$ MeVfm$^{2}$ and 21 sets have $g_{\rm sur}\approx 18$ MeVfm$^{2}$. For the three sets of parameters of the ImQMD model listed in Table 1, the surface energy coefficient $g_{\rm sur}$ varies in a range of $18 \pm 1.5$ MeVfm$^2$ which is generally consistent with the results of most Skyrme parameter sets.  Since the surface coefficient is different for the three sets of parameters, it could be helpful to obtain the information on this parameter from different heavy-ion reactions.

\begin{figure}
\includegraphics[angle=0,width=0.7\textwidth]{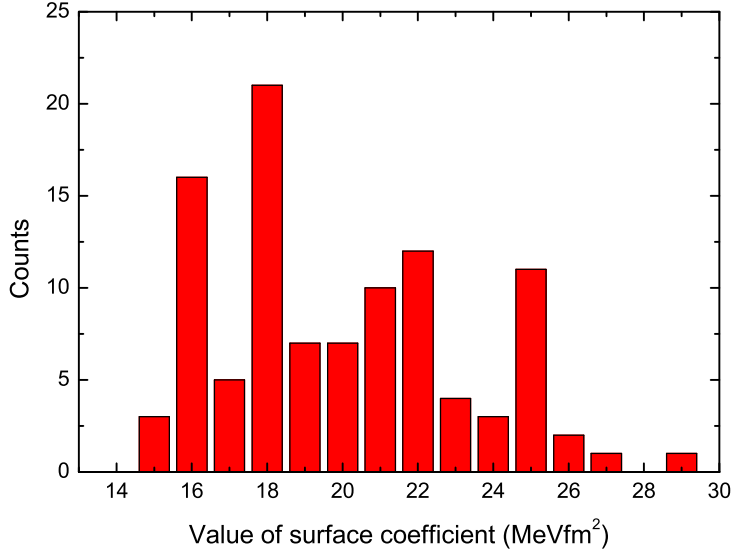}
\caption{(Color online) Distribution of the values of the surface coefficients $g_{\rm sur}$ from 103 sets of Skyrme forces. }
\end{figure}

\begin{table}
 \caption{ Model parameters adopted in this work.}
\begin{tabular}{lccccccccccc}
\hline Parameter & $\alpha $ & $\beta $ & $\gamma $ &$%
g_{\rm sur}$ & $ g_{\tau }$ & $\eta $ & $C_{s}$ & $\kappa _{s}$ &
$\rho
_{0}$ & ~~$\sigma_0$~~ & ~~$\sigma_1$~~ \\
 & (MeV) & (MeV) &  & (MeVfm$^{2}$) & (MeV) &  & (MeV) & (fm$^{2}$) &
 (fm$^{-3}$) & (fm) & (fm) \\ \hline
 SkP* & $-356$ & 303 & 7/6 & 19.5 & 13 & 2/3 &  35  & 0.65  & 0.162 & 0.94 & 0.018\\
IQ3a & $-207$ & 138 & 7/6 & 16.5 & 14 & 5/3 &  34  & 0.4  & 0.165 & 0.94 & 0.020\\
IQ3b & $-207$ & 138 & 7/6 & 18.0 & 14 & 5/3 &  34  & 0.6  & 0.165 & 0.94 & 0.018\\
  \hline
\end{tabular}
\end{table}

\begin{center}
\textbf{B. Initialization and stability of nuclei  }
\end{center}

In this work, the neutron skin thickness of neutron-rich nuclei is taken into account in the initialization of the ImQMD model. Based on the 4-parameter nuclear charge radii formula proposed in Ref. \cite{Radii}
\begin{eqnarray}
R_c= 1.226 A^{1/3} + 2.86 A^{-2/3} -1.09 (I-I^2) + 0.99 \Delta E/A,
\end{eqnarray}
with which the 885 measured charge radii can be reproduced with a rms deviation of 0.022 fm, and the linear relationship between the neutron skin thickness $\Delta R_{np}=\langle r_n ^2\rangle ^{1/2}-\langle r_p ^2\rangle ^{1/2}$ and the isospin asymmetry $I=(N-Z)/A$ \cite{Trz01}
\begin{eqnarray}
\Delta R_{np}=0.9 I -0.03,
\end{eqnarray}
one can obtain the proton radii $R_p=\sqrt{\frac{5}{3}  \left [\langle r_c ^2\rangle  -0.64 \right ]}$ from the root-mean-squre (rms) charge radii $\langle r_c ^2\rangle^{1/2}=\sqrt{\frac{3}{5}} R_c$ and the neutron radii $R_n=\sqrt{\frac{5}{3}} \left [\langle r_p ^2\rangle^{1/2}  + \Delta R_{np} \right ]$. The nucleon positions are sampled within the hard sphere with a radius $R_p -w_r$ for the protons and $R_n- w_r$ for the neutrons, respectively. Here, $w_r=0.8$ fm is to take into account the influence of the width of the wave-packet in the coordinate space. In Fig. 2, we show the density distribution of $^{208}$Pb and $^{132}$Sn. The scattered symbols denote the results of the ImQMD model at the initial time, and the curves denote the corresponding results of the Skyrme Hartree-Fock calculations with the force SkM* \cite{skm}. We find that although the nucleon positions are sampled within the hard sphere, the neutron skin thickness and the surface diffuseness of nuclei at the initial time can be reasonably well described due to the gaussian wave-packet for each nucleon in the ImQMD model. Here, we would like to emphasize that the density distribution of nuclei in the realistic calculations for the heavy-ion reactions could be different from the initial density distributions given by the sampling. The density distribution averaged over times during the evolution at their ground state for few thousands fm/c which reflects the self-consistent effect of the mean field \cite{Giu14} should be further checked.

\begin{figure}
\includegraphics[angle=0,width=0.8\textwidth]{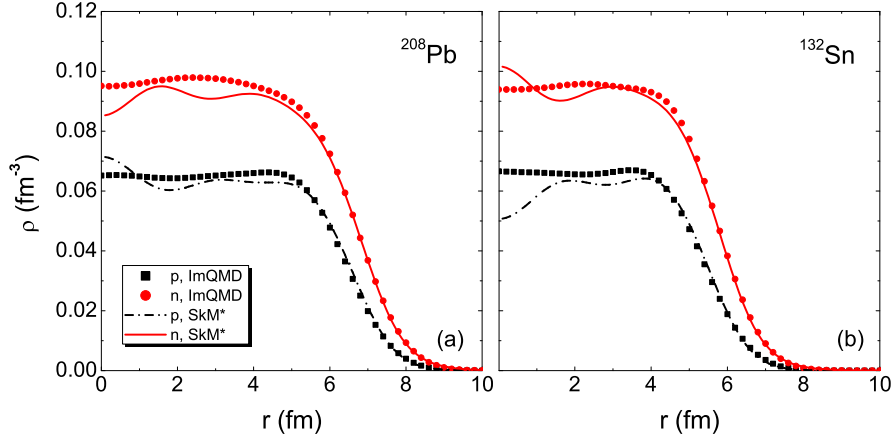}
\caption{(Color online) Density distribution of $^{132}$Sn and $^{208}$Pb at the initial time. The red for neutrons and the black for protons.}
\end{figure}

\begin{figure}
\includegraphics[angle=0,width=0.9\textwidth]{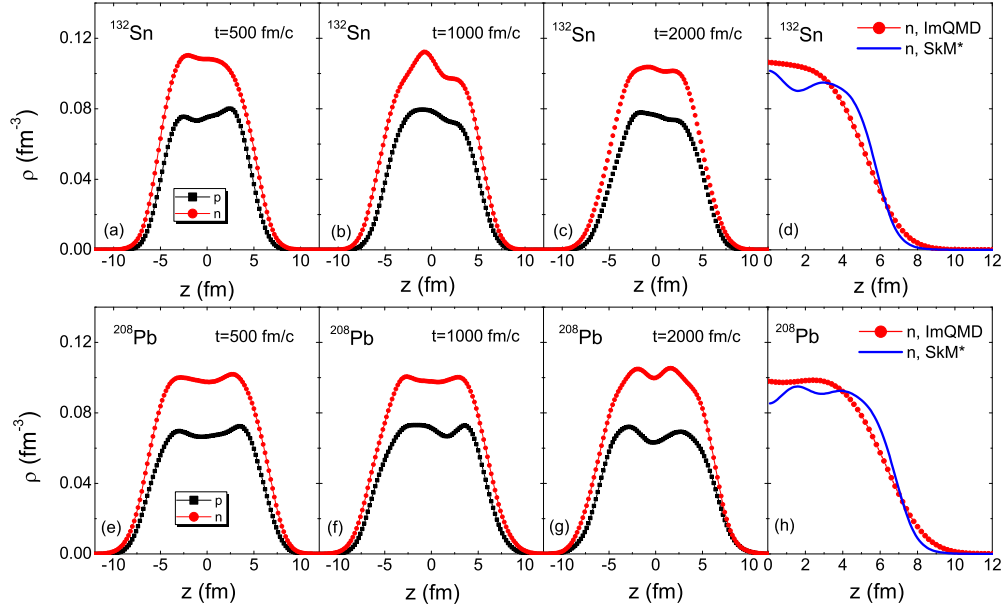}
\caption{(Color online) Density distribution of individual nucleus $^{132}$Sn and $^{208}$Pb during the time evolution. Here, we create 100 events for each nucleus. In (d) and (h), the solid circles denote the neutron density of $^{132}$Sn and $^{208}$Pb averaged over times in the ImQMD simulations. The solid curves denote the neutron density from the Skyrme Hartree-Fock calculations with the force SkM*.   }
\end{figure}

With the sampled nucleon positions, the nuclear potential energy of the nucleus can be calculated with Eq.(4). The momentum of the $i$-th nucleon is then sampled within the local Fermi sphere with a radius $ \hbar [3\pi^2 \rho_q ({\bf r_i}) ]^{1/3}-w_p$, where $q=n$ for neutrons and $q=p$ for protons. $w_p$ is to consider the influence of the width of the wave-packet in the momentum space and its value is determined by the experimental binding energy $BE$ (in negative value) of the sampled nuclei. If the ground state energy of the nuclei calculated with Eq.(3) falls into the range of $BE \pm 0.05$ MeV and at the same time the distance between any two nucleons in the phase space $|{\bf r_i}-{\bf r_j}| \cdot |{\bf p_i}-{\bf p_j}|\ge 255$ fm$\cdot$MeV/c, the sampled nuclei will be used in the simulations.

For the realistic simulations of heavy-ion fusion reactions at energies around the Coulomb barrier, it generally takes several hundreds fm/c to about 1000 fm/c for the projectile nuclei to approach the target nuclei and to overcome the Coulomb barrier. By adopting a proper force such as IQ3 for the description of the mean field, we find that most sampled nuclei along the stability line can remain stable for thousands fm/c and the number of spurious nucleon emission is very small. We have checked that the average numbers of spurious emitted nucleons at $t=2000$ fm/c are merely 1.1 for the individual $^{92}$Zr nuclei and 2.6 for $^{132}$Sn with the parameter set IQ3a, respectively. In Fig. 3, we show the density distribution of $^{208}$Pb and $^{132}$Sn by adopting IQ3a. One sees that the central density of the sampled nuclei remains about 0.17 fm$^{-3}$ for $^{208}$Pb and 0.18 fm$^{-3}$ for $^{132}$Sn during the whole evolution of 2000 fm/c, although the surface diffuseness of nuclei is a little larger than that at the initial time. In Fig.3 (d) and (h), we show the corresponding density distribution of neutrons averaged over times which reflects the self-consistent effect of the mean field used in the calculations. From Fig. 3(d) and (h), one sees that the surface diffuseness of neutron distribution is a little larger than the results of the Skyrme Hartree-Fock calculations with the force SkM*. In addition, we note that the total energies of the systems are reasonably well conserved during the time evolution of 2000 fm/c, even the momenta of some nucleons change abruptly and occasionally due to the spurious "elastic scattering" between nucleons required in the Fermi constraint. These calculations imply that the reaction systems simulated with the ImQMD model can remain stable for a long enough period of time during the capture process of the heavy-ion fusion reactions, which is also of crucial importance for the reliable description of the fragment yields and the reaction cross sections in heavy-ion collisions at intermediate energies.

\begin{center}
\textbf{III. RESULTS AND DISCUSSIONS}
\end{center}

In this section, we first systematically calculate the fusion excitation functions of heavy-ion fusion reactions $^{16}$O+$^{76}$Ge, $^{16}$O+$^{154}$Sm, $^{40}$Ca+$^{96}$Zr and $^{132}$Sn+$^{40}$Ca. Then, we study the influence of surface-symmetry term and the dynamical evolution of the density at neck region in neutron-rich reaction systems. Simultaneously, the validity of the model for describing fusion between stable nuclei will be further tested.

 \newpage

\begin{center}
\textbf{A. Fusion cross sections}
\end{center}

\begin{figure}
\includegraphics[angle=0,width=1 \textwidth]{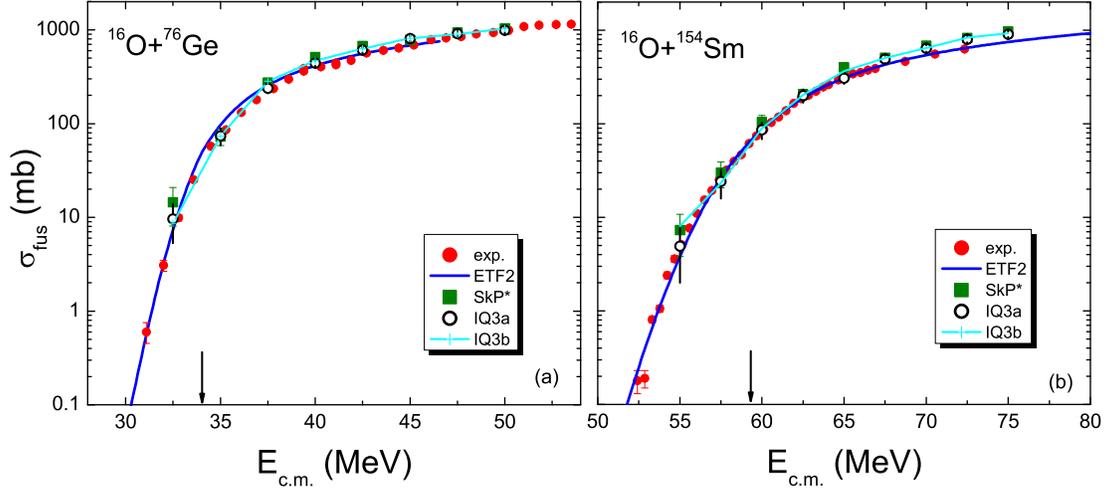}
\caption{(Color online)  Fusion excitation functions of $^{16}$O+$^{76}$Ge and $^{16}$O+$^{154}$Sm. The solid circles denote the experimental data taken from \cite{OGe} and \cite{OSm}, respectively. The blue curves denote the results with an empirical barrier distribution in which the fusion barrier is obtained by using the Skyrme energy-density functional together with the extended Thomas-Fermi (ETF2) approximation \cite{liumin06,Wang09}. The solid squares and open circles denote the results of ImQMD with the parameter set SkP* and IQ3a, respectively. The statistical errors in the ImQMD calculations are given by the error bars. The arrows denote the most probable barrier height based on the barrier distribution function adopted in the ETF2 approach. }
\end{figure}

It is known that the fusion potential between two nuclei is closely related to the surface properties of the nuclei. The neutron skin thickness of neutron-rich nuclei in heavy-ion fusion reactions should affect the fusion barrier and thus the fusion cross sections. To explore the influence of the symmetry potential on the fusion excitation function, the fusion cross sections of $^{16}$O+$^{76}$Ge, $^{16}$O+$^{154}$Sm, $^{40}$Ca+$^{96}$Zr and $^{132}$Sn+$^{40}$Ca are systematically investigated
with the ImQMD model by adopting different parameter sets. Through creating certain bombarding events (about 100 to 200) at each incident energy $E_{\rm c.m.}$ and each impact parameter $b$, and counting the number of fusion events, we obtain the fusion probability (or capture probability for reactions leading to super-heavy nuclei) $g_{\rm fus}(E_{\rm c.m.},b)$ of the reaction, by which the fusion (capture) cross
section can be calculated:
\begin{equation}
\sigma _{\rm fus}(E_{\rm c.m.})=2\pi \int b \, g_{\rm fus} \, db
\simeq 2\pi \sum b \, g_{\rm fus} \, \Delta b.
\end{equation}

\begin{figure}
\includegraphics[angle=0,width=1 \textwidth]{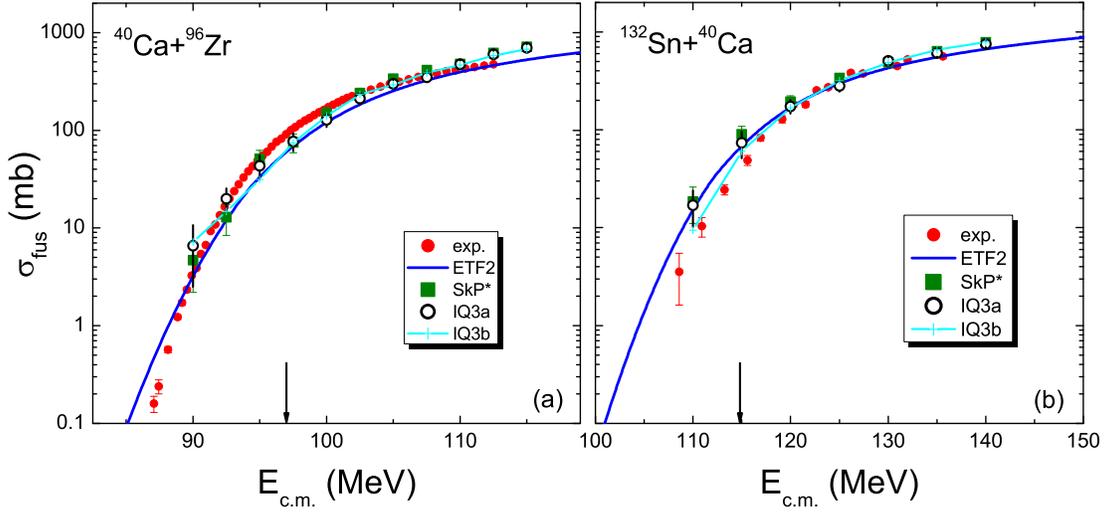}
\caption{(Color online) The same as Fig.4 but for $^{40}$Ca+$^{96}$Zr \cite{Timm} and $^{132}$Sn+$^{40}$Ca \cite{CaSn}.}
\end{figure}

To consider the influence of the Coulomb excitation, the initial distance $R_0$ between the projectile and target should be much larger than the fusion radius. The collective boost to the sampled initial nuclei is given by $E_{\rm kin}=E_{\rm c.m.}-Z_1 Z_2 e^2/R_0$ at the initial time, with the center-of-mass energy $E_{\rm c.m.}$, the charge number $Z_1$ and $Z_2$ for the projectile and target nuclei, respectively. Here, the initial distance between the reaction partners at z-direction (beam direction) is taken to be $d_0=30$ fm for the intermediate reaction systems and 40 fm for the ones with stronger Coulomb repulsion such as $^{132}$Sn+$^{40}$Ca. In the calculation of the fusion (capture) probability, the time evolution of a certain simulating event will be terminated to save the CPU time and the event will be counted as a fusion (capture) event if the center-to-center distance between the two nuclei is smaller than the nuclear radius of the compound nuclei (which is much smaller than the fusion radius).

\begin{figure}
\includegraphics[angle=0,width=0.9 \textwidth]{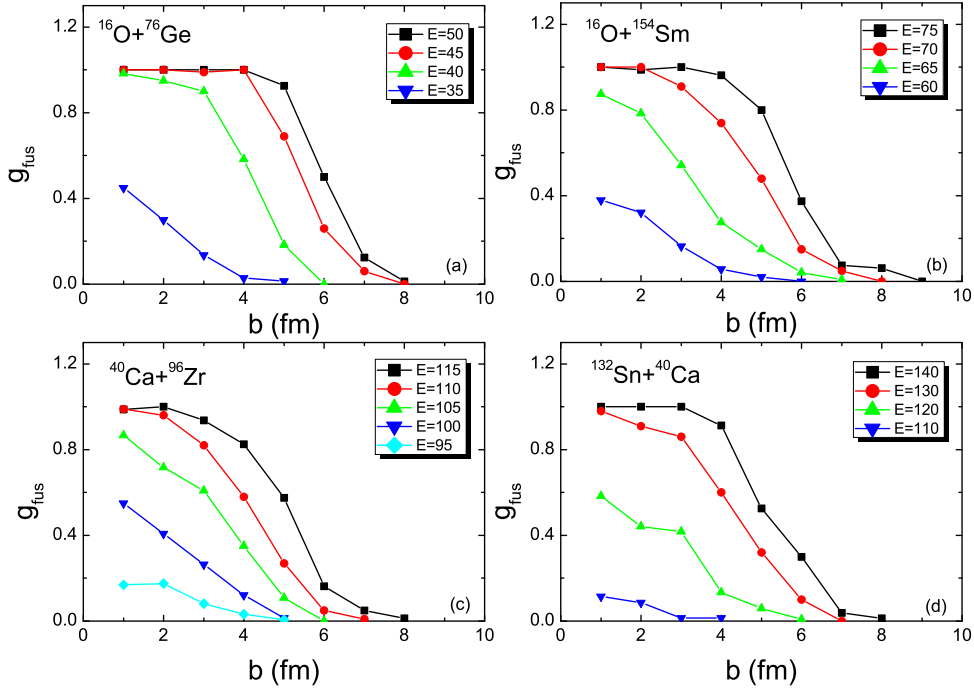}
\caption{(Color online) Fusion probability of the fusion reactions with IQ3a as a function of impact parameter $b$.}
\end{figure}

Fig. 4 shows the comparison of the calculated results and the experimental data for the fusion reactions $^{16}$O+$^{76}$Ge and $^{16}$O+$^{154}$Sm. The results of $^{40}$Ca+$^{96}$Zr and $^{132}$Sn+$^{40}$Ca are shown in Fig. 5. The experimental data can be reasonably well reproduced with all the three sets of parameters listed in Table 1.
The arrows denote the most probable barrier height based on the barrier distribution function adopted in the ETF2 approach \cite{liumin06}. For fusion at energies below the Coulomb barrier, the dynamical fluctuations in the densities of the reaction partners result in the lowering of the barrier height for the fusion events, which will be further discussed later. For $^{132}$Sn+$^{40}$Ca, the fusion cross sections at sub-barrier energies are slightly over-predicted by both the ETF2 and the ImQMD calculations, which is probably due to the influence of the shell effect in the doubly-magic nucleus $^{132}$Sn. In the semi-classical ImQMD model, the shell effect is not self-consistently considered in the simulations. In Fig. 6, we show the calculated fusion probability as a function of impact parameter. For the fusion reactions at energies above the Coulomb barrier, the fusion probability looks like a Fermi distribution, i.e. the fusion probability is about one for the central and mid-central collisions. At energies around the Coulomb barrier, the fusion probability decreases quickly with the impact parameter, which implies that the centrifugal potential due to the angular momentum affects the results significantly at this energy region.

\begin{figure}
\includegraphics[angle=0,width=1.0 \textwidth]{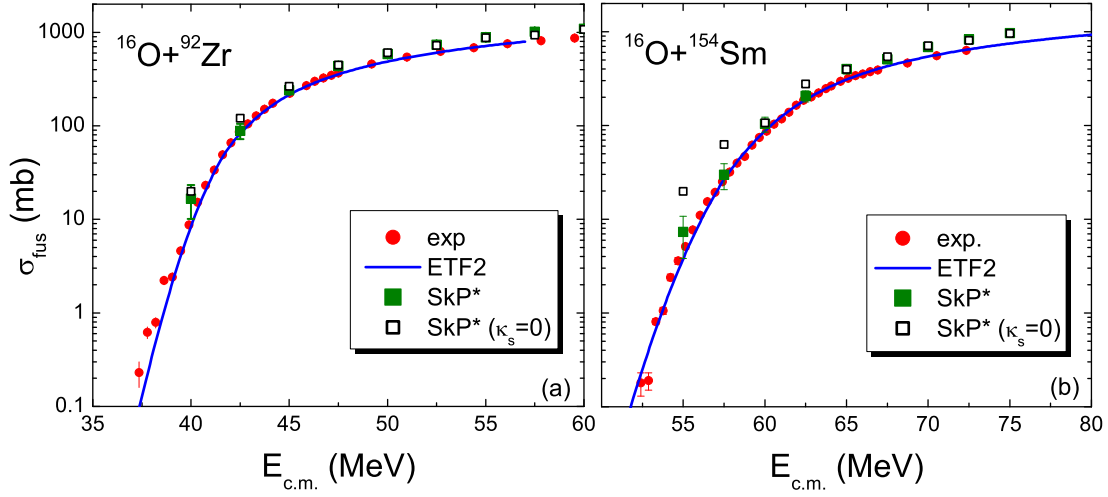}
\caption{(Color online) Fusion excitation functions of $^{16}$O+$^{92}$Zr \cite{OZr} and $^{16}$O+$^{154}$Sm.  The open squares denotes the results with SkP* but setting the surface-symmetry coefficient $\kappa_s=0$. }
\end{figure}

\begin{center}
\textbf{B. Influence of surface-symmetry term and neutron density at neck}
\end{center}

In the macroscopic-microscopic mass formula \cite{Wang}, the surface-symmetry term plays an important role on the description of the symmetry energy coefficient of finite nuclei. In this work, we simultaneously investigate the influence of surface-symmetry coefficient $\kappa_s$ on the fusion cross sections of heavy-ion fusion reactions from the microscopic dynamics simulations. In Fig. 7, we compare the fusion excitation functions of $^{16}$O+$^{92}$Zr \cite{OZr} and $^{16}$O+$^{154}$Sm based on the ImQMD calculations by using the parameter set SkP*, with and without the surface-symmetry term being taken into account. For the reactions with neutron-rich nuclei such as $^{16}$O+$^{154}$Sm, the influence of surface symmetry term is much more obvious than that in $^{16}$O+$^{92}$Zr. The fusion cross sections $^{16}$O+$^{154}$Sm at sub-barrier energies are significantly suppressed by the
surface-symmetry term. We also note that the surface-symmetry term is helpful to improve the stability of the sampled neutron-rich nuclei in the initialization of the ImQMD model.

\begin{figure}
\includegraphics[angle=0,width=0.8 \textwidth]{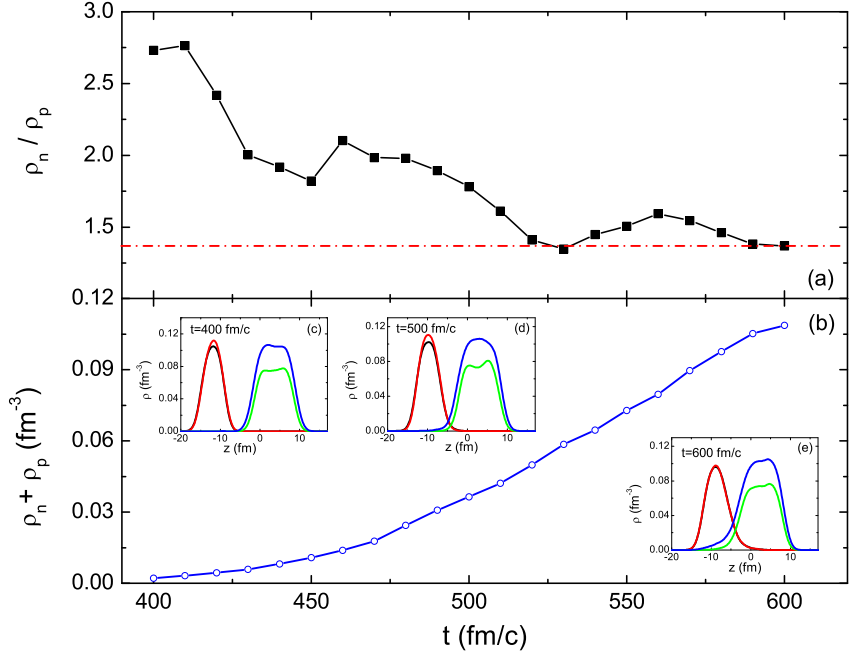}
\caption{(Color online) Time evolution of the density at neck region in fusion reaction $^{132}$Sn+$^{40}$Ca at an incident energy of $E_{\rm c.m.}=115$ MeV. (a) Ratio of neutron-to-proton density at neck. The dash-dotted line denotes the corresponding ratio of the compound nuclei. (b) Density of the fusion system at neck. The sub-figures show the density distribution of the fusion system at $t=400, 500$ and 600 fm/c, with the red and the blue curves for the neutrons and the others for the protons.}
\end{figure}

In Fig. 8, we show the time evolution of the density at neck region in the fusion reaction $^{132}$Sn+$^{40}$Ca at an incident energy of $E_{\rm c.m.}=115$ MeV. Here, the density distribution of 328 fusion events over a total of 1000 simulation events is studied for the head-on collisions with the parameter set IQ3a. At $t=400$ fm/c, the reaction partners begin to touch each other, and the ratio of neutron-to-proton density $\rho_n/\rho_p$ reaches 2.7 which is higher than the $N/Z$ of the compound nucleus by a factor of two. With the increase of the density at neck, the value of $\rho_n/\rho_p$ deceases with some oscillations and gradually approaches the corresponding neutron-to-proton ratio of the compound nucleus ($N/Z=1.37$). The extremely neutron-rich density at the neck region can significantly suppress the height of the Coulomb barrier for the fusion reactions induced by neutron-rich nuclei at energies around and below the barrier. Furthermore, one can see from Fig.8(e) that the surface diffuseness of the reaction partners at the neck side is obviously larger than that at the other side due to the transfer of nucleons. The energy dependence of the fusion potential due to the dynamical density evolution can be clearly observed from the ImQMD \cite{ImQMD2010,ImQMD2012} and TDHF \cite{Wash08,Umar14} simulations.

\begin{center}
\textbf{C. Fusion between stable nuclei and Sub-barrier fusion}
\end{center}

\begin{figure}
\includegraphics[angle=0,width=1 \textwidth]{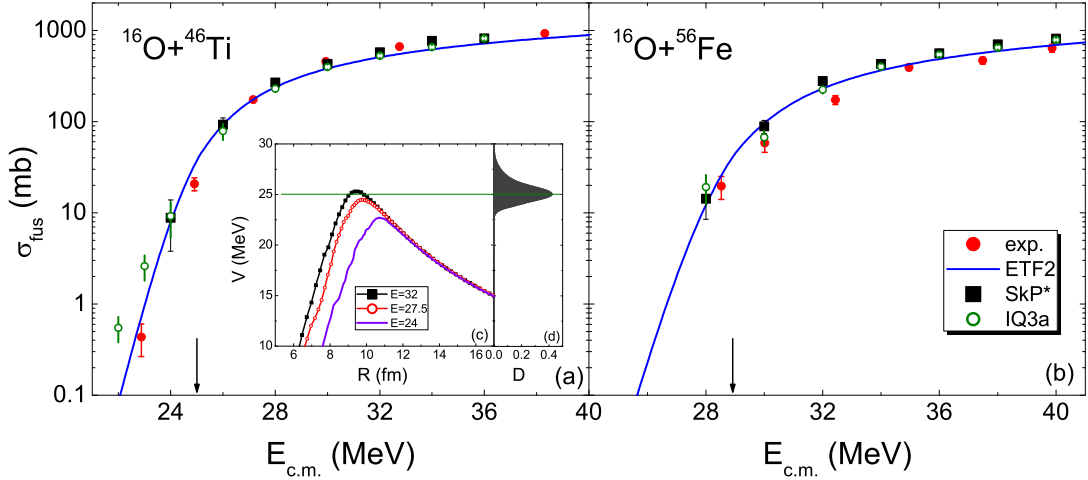}
\caption{(Color online) Fusion excitation functions of $^{16}$O+$^{46}$Ti \cite{Lig90} and $^{16}$O+$^{56}$Fe \cite{Fun93}. (c) Dynamical nucleus-nucleus potential of $^{16}$O+$^{46}$Ti from the head-on collisions of ImQMD simulations at three different incident energies with the parameter set SkP*. (d) Barrier distribution function of $^{16}$O+$^{46}$Ti adopted in the ETF2 calculations. }
\end{figure}

\begin{figure}
\includegraphics[angle=0,width=1 \textwidth]{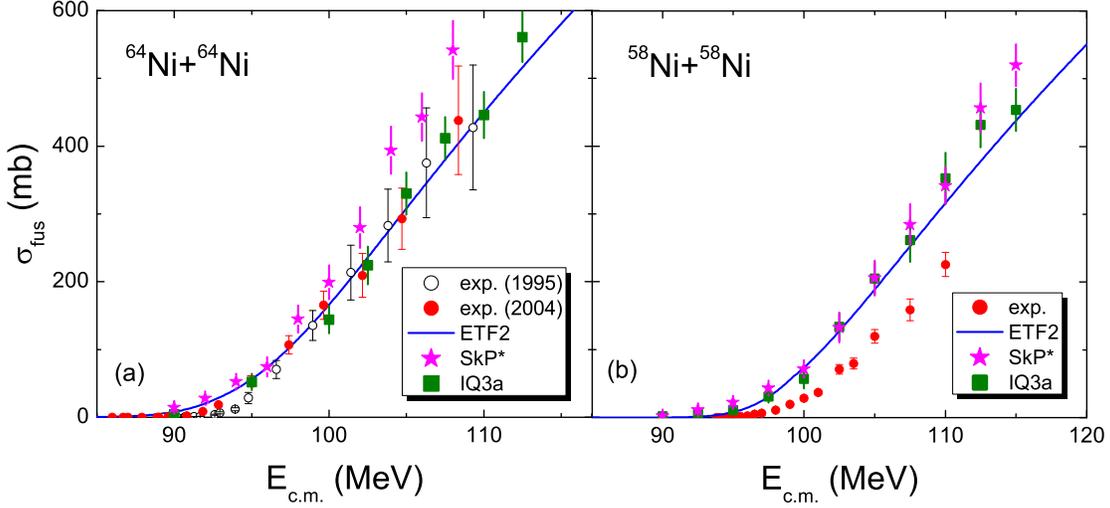}
\caption{(Color online) Fusion excitation functions of $^{64}$Ni+$^{64}$Ni \cite{Ack95,Jiang04} and $^{58}$Ni+$^{58}$Ni \cite{Beck81}. The stars and squares denote the results from the ImQMD model with the parameter sets SkP* and IQ3a, respectively.  }
\end{figure}

To further test the validity of the model for the description of fusion between stable nuclei and to explain the sub-barrier fusion phenomena, the fusion cross sections of $^{16}$O+$^{46}$Ti, $^{16}$O+$^{56}$Fe and $^{58}$Ni+$^{58}$Ni are also studied with the ImQMD model. Fig. 9 shows the fusion excitation functions of $^{16}$O+$^{46}$Ti and $^{16}$O+$^{56}$Fe. At energies around the Coulomb barrier, the experimental data can be well reproduced with both the ETF2 and the ImQMD calculations. At deep sub-barrier energies, the experimental data for $^{16}$O+$^{46}$Ti are over-predicted by the ImQMD calculations, which is probably due to the over-predicted nuclear surface diffuseness in the ImQMD simulations (see Fig. 3). Here, we create 20000 events to simulate the fusion reaction $^{16}$O+$^{46}$Ti at a deep sub-barrier energy of $E_{\rm c.m.}=22$ MeV. To illustrate the sub-barrier fusion, in Fig. 9(c) we show the dynamical nucleus-nucleus potential \cite{ImQMD2010,ImQMD2012} of $^{16}$O+$^{46}$Ti by using the ImQMD model at three different incident energies with the parameter set SkP*.  The average barrier heights for the fusion events are 25.3, 24.5 and 22.7 MeV at the incident energy of 32, 27.5 and 24 MeV, respectively. Fig. 9(d) shows the barrier distribution function of this fusion reaction adopted in the ETF2 calculations \cite{liumin06}. The most probable barrier height is 25.0 MeV. For fusion at the energy of $E_{\rm c.m.}=24$ MeV which is lower than the most probable barrier height, the realistic dynamical barrier height calculated from the fusion events in the simulations is slightly lower than the incident energy due to the dynamical fluctuations of the densities and nucleon transfer. From the point of view of the semi-classical ImQMD model based on event-by-event simulations, the "sub-barrier" fusion is a process that the rare projectile nuclei surmount rather than tunnel through the suppressed potential barrier. In Fig. 10, we compare the fusion cross sections of $^{64}$Ni+$^{64}$Ni and $^{58}$Ni+$^{58}$Ni at energies above the Coulomb barrier. It is found that the shell and isospin effects mainly influence the fusion cross sections near and below the Coulomb barrier \cite{liumin06}. At energy above the barrier, the fusion cross sections can be approximately described by the formula $\sigma_{\rm fus}\approx \pi R_{\rm fus}^2 (1- B/ E_{\rm c.m.})$, with the fusion radii $R_{\rm fus}$ and the barrier height $B$.  One sees that for the neutron-rich fusion reaction $^{64}$Ni+$^{64}$Ni, the experimental data at energies above the Coulomb barrier can be reasonably well described by both ETF2 and ImQMD calculations, whereas the same theoretical approaches fail to reproduce the experimental data for $^{58}$Ni+$^{58}$Ni and the reason is not clear.  For $^{64}$Ni+$^{64}$Ni, the results of SkP* are slightly larger than those of IQ3a due to the relatively larger surface coefficient adopted. However, for $^{58}$Ni+$^{58}$Ni the results of SkP* and IQ3a are comparable. More precise experimental measurements for the fusion and quasi-elastic scattering cross sections of $^{58}$Ni+$^{58}$Ni, and the structure properties of $^{58}$Ni could be helpful for extracting the information of the potential barrier and understanding the deviations.

\begin{center}
\textbf{IV. SUMMARY}
\end{center}

In summary, the heavy-ion fusion reactions induced by neutron-rich nuclei have been investigated by using the improved quantum molecular dynamics model with three sets of parameters. With the neutron skin thickness of nuclei being taken into account in the initialization of the model, the stability of the individual nuclei have been checked and the fusion excitation functions of heavy-ion fusion reactions $^{16}$O+$^{76}$Ge, $^{16}$O+$^{92}$Zr, $^{16}$O+$^{154}$Sm, $^{40}$Ca+$^{96}$Zr and $^{132}$Sn+$^{40}$Ca at energies around the Coulomb barrier can be reasonably well reproduced. The slope parameter of the nuclear symmetry energy adopted in the calculations is about $L \approx 78$ MeV and the surface coefficient is $g_{\rm sur}=18\pm 1.5$ MeVfm$^2$. We also note that the surface-symmetry term significantly influences the stability of neutron-rich nuclei and the fusion cross sections of neutron-rich fusion systems at energies below the Coulomb barrier. The extremely large ratio of the neutron-to-proton density at the neck region and the enhanced surface diffuseness at the neck side due to the microscopic dynamical evolution can significantly suppress the barrier height for the fusion reactions induced by the neutron-rich nuclei at energies around and below the Coulomb barrier. The validity of the model for description of fusion between stable nuclei has also been tested. At energies around the Coulomb barrier, the measured fusion cross sections of $^{16}$O+$^{46}$Ti and $^{16}$O+$^{56}$Fe can be well reproduced with both ETF2 and ImQMD calculations, whereas the experimental data of $^{58}$Ni+$^{58}$Ni are significantly over-predicted by the two theoretical approaches, even at energies above the Coulomb barrier.

\begin{center}
\textbf{ACKNOWLEDGEMENTS}
\end{center}
This work was supported by National Natural Science Foundation of
China (Nos 11275052, 11365005, 11365004, 91126010, 11375062) and National Key Basic Research Program of China (Grant No. 2013CB834400).

\end{document}